# DataPro – A Standardized Data Understanding and Processing Procedure: A Case Study of an Eco-driving Project

Zhipeng Ma[1[0000-0002-4049-539X]], Bo Nørregaard Jørgensen[1[0000-0001-5678-6602]] and Zheng Grace Ma[1[0000-0002-9134-1032]]

[1] SDU Center for Energy Informatics, Maersk Mc-Kinney Moller Institute, The Faculty of Engineering, University of Southern Denmark, DK-5230 Odense, Denmark
`{zhma, bnj, zma}@mmmi.sdu.dk`

**Abstract.** A systematic pipeline for data processing and knowledge discovery is essential to extracting knowledge from big data and making recommendations for operational decision-making. The CRISP-DM model is the de-facto standard for developing data-mining projects in practice. However, advancements in data processing technologies require enhancements to this framework. This paper presents the DataPro (a standardized data understanding and processing procedure) model, which extends CRISP-DM and emphasizes the link between data scientists and stakeholders by adding the "technical understanding" and "implementation" phases. Firstly, the "technical understanding" phase aligns business demands with technical requirements, ensuring the technical team's accurate comprehension of business goals. Next, the "implementation" phase focuses on the practical application of developed data science models, ensuring theoretical models are effectively applied in business contexts. Furthermore, clearly defining roles and responsibilities in each phase enhances management and communication among all participants. Afterward, a case study on an eco-driving data science project for fuel efficiency analysis in the Danish public transportation sector illustrates the application of the DataPro model. By following the proposed framework, the project identified key business objectives, translated them into technical requirements, and developed models that provided actionable insights for reducing fuel consumption. Finally, the model is evaluated qualitatively, demonstrating its superiority over other data science procedures.

## 1  Introduction

The increasing data proliferation within industrial digitalization has impacted various industrial sectors by emerging technologies in the area of the Internet of Things (IoT) and Cyber-physical Systems (CPS), enhancing the ability to gather and share industrial big data [1-2]. Therefore, the systematic application of data science techniques to extract knowledge from data becomes increasingly popular in transportation, manufacturing, and so on, enabling more informed recommendations and decision-making to optimize operational efficiency [3].

Among several employed methodologies, the Cross Industry Standard Process for Data Mining (CRISP-DM) is preferred by the majority of stakeholders [4]. The CRISP-DM model, funded by the European Union, was presented in 1999 and published as a step-by-step application guide [5]. It afterward becomes the "de-facto standard for developing data mining and knowledge discovery projects" [6]. This model provides a clear and structured framework for navigating through various stages of a data science project. Moreover, its general applicability ensures that it can be adapted to different application sectors and diverse data science tasks, making it a convincing choice for stakeholders seeking effective project management and implementation.

The CRISP-DM model has been proposed for over two decades and is still selected by about half of the practitioners as the baseline framework for executing data science projects [4]. The methodologies to derive data values have exponentially increased in recent years due to the development of both hardware and software. Firstly, the improvements in sensors and storage devices have enhanced the capabilities of data collection and storage. Secondly, the rapid advancement of machine learning and other artificial intelligence technologies has provided more accurate and efficient tools for decision-making and data analytics.

To align the CRISP-DM model with state-of-the-art data science tasks and specific domains, many studies have refined it by augmenting phases and enriching execution activities within the phases. The model developed in [7] combines the CRISP-DM and the agile project management method to analyze the end-to-end supply chain, handling the organizational and socio-technical challenges of big data analytics in supply chain management. A phase named "Operation and Maintenance" is added in [8] to embed a task-based framework in the manufacturing sector for associating tasks with skills. The QM-CRISP-DM model proposed in [9] incorporates process improvement and error analysis in production quality management. Each phase in the model is extended by integrating Six Sigma tools to provide a concrete development toolbox.

Despite ongoing refinements to the CRISP-DM model, significant opportunities remain to address its limitations and challenges. Firstly, this model lacks a communication mechanism with the team members and stakeholders to determine when to loop back to a previous phase [10]. This deficiency can lead to misalignment between project expectations and outcomes, as stakeholders might be unaware of the need for adjustments or refinements during the project's lifecycle. Secondly, the framework does not clearly define the roles and responsibilities of stakeholders, leading to ambiguity in their involvement. This gap limits the value transmission between data science and business domains, as it hampers the integration of domain expertise into the data science process and restricts feedback loops that are critical for ensuring that the analysis aligns with business objectives.

To address the aforementioned challenges, this paper proposes DataPro, an enriched data understanding and processing framework for data science projects. Based on CRISP-DM, DataPro emphasizes the connection between data scientists and stakeholders, adding two phases: "technical understating" and "implementation". The "technical understanding" phase bridges the gap between business demands and the corresponding technical requirements and constraints, ensuring that the technical team clearly and precisely understands business goals. The "implementation" phase focuses on the practical application of the developed data science models, which ensures that theoretical models are effectively translated into actionable solutions. Furthermore, the roles and responsibilities of different participants are clearly defined and emphasized in each phase for better management and communication.

Subsequently, a case study on an energy-related data science project in the transportation sector demonstrates the practical application and effectiveness of the DataPro model in real-world scenarios. This project aimed to reduce fuel consumption and greenhouse gas emissions in Danish public transport. Using the DataPro framework, it identified business goals, translated them into technical requirements, and developed models to reduce fuel consumption. Key findings included four fuel efficiency groups and the impact of driving behavior and route characteristics.

Finally, the model is evaluated using four qualitative criteria, showcasing its superiority over other data science procedures.

The remainder of the paper is organized as follows. Section 2 introduces the CRISP-DM model and analyzes its strengths and drawbacks. Section 3 outlines the principle and details of the proposed DataPro model. Section 4 shows the application in a case study. The proposed model is evaluated in Section 5, and Section 6 discusses and concludes the present work.

## 2    Related Work

### 2.1    CRISP-DM

The CRISP-DM model is a widely adopted industry-oriented framework for implementing the generic knowledge discovery process, converting business challenges into data science activities [11]. It consists of six phases: including business understanding, data understanding, data preparation, modeling, evaluation, and deployment. Fig. 1 illustrates the workflow for executing the CRISP-DM model and the interrelationship among its phases.

**Business Understanding.** In the business understanding phase, the primary focus is on determining the business objectives. It is crucial for data scientists to gain a thorough insight into these goals as they pertain to data analytic efforts. During this stage, stakeholders play a key role in providing background information about the current business situation. They are responsible for documenting specific business objectives and determining the evaluation criteria from a business perspective.

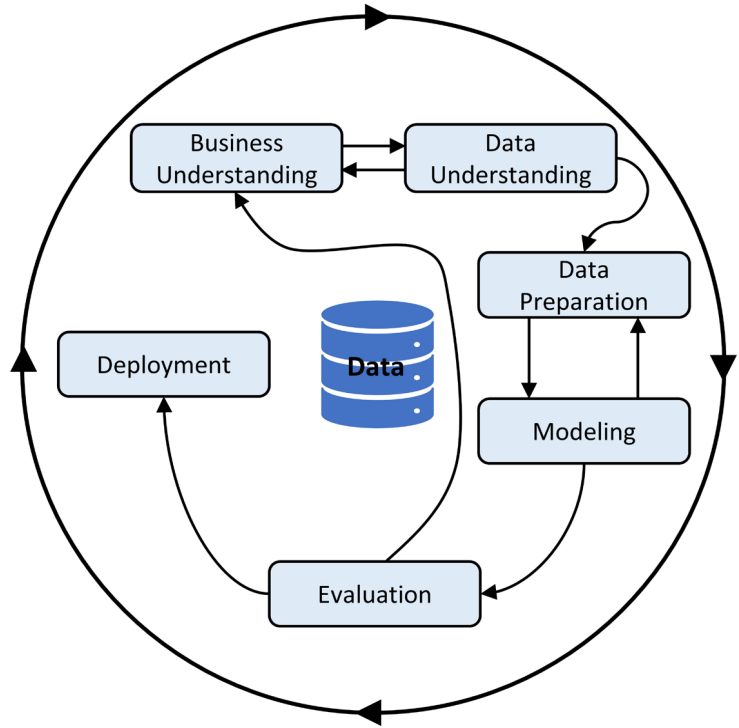

**Fig. 1.** The CRISP-DM process model [5].

**Data Understanding.** In the data understanding phase, the collected data is explored to uncover hidden information and insights. This exploration is essential for understanding the underlying patterns and characteristics of the data, which can significantly impact the outcomes of subsequent data preparation. Additionally, the data quality is examined and the plan to address the identified quality issues is developed in this stage. By thoroughly assessing the data quality, data scientists can identify and address potential issues such as missing values, inconsistencies, and inaccuracies that could lead to unexpected problems in later stages [12].

There is a feedback loop between data understanding and business understanding, as the insights gained during data understanding lead to a reevaluation of the initial business objectives. For instance, discovering unexpected patterns or trends in the data might prompt stakeholders to redefine the scope of the project.

**Data Preparation.** Data preparation is one of the most important and time-consuming phases in a data science project, as the quality and structure of the data significantly influence the accuracy and trustworthiness of the model in the next phase. Data preparation typically involves several key tasks, including data cleaning, data transformation, data integration, and other necessary data preprocessing strategies. After executing this phase, data scientists are capable of creating robust and reliable datasets that form the foundation of accurate and insightful analytics.

**Modeling.** In the modeling phase, the data analytic models are selected and the parameters are fine-tuned to process the prepared datasets. This stage involves choosing the appropriate modeling techniques and adjusting them to achieve optimal performance based on the specific characteristics of the data and the business objectives.

It is common for data scientists to revert to the data preparation phase during modeling. This iteration may be necessary to perform additional manipulations or preprocessing steps required by the chosen models. These adjustments can include further cleaning, transformation, or feature engineering to better align the data with the model's requirements.

**Evaluation.** In the evaluation phase, the results generated from the modeling phase are assessed against the data processing goals and the business objectives established earlier in the project. Firstly, the performance of the models is evaluated using various metrics and validation techniques. This assessment helps determine how well the models perform on the prepared datasets and whether they meet the expected standards. Secondly, the results are compared with business objectives to ensure that the models not only perform well technically but also deliver actionable insights and value from a business perspective.

**Deployment.** In the deployment phase, the results are integrated into the business process. During this phase, the final report is written to document the detailed steps required to execute the project. Additionally, a comprehensive project review is conducted to evaluate the project outcomes, insights gained, and opportunities for enhancement, ensuring that future projects benefit from the experiences and knowledge acquired.

## 2.2 Strengths

The CRISP-DM model defines a set of intuitive steps for the team to follow, which are straightforward to understand [10]. Data scientists naturally adhere to a similar process without the project management direction. Therefore, they can easily identify the CRISP-DM phases and perform iterations when working on a project. Furthermore, this model aligns the technical work with business objectives, ensuring that data scientists properly comprehend the business problem through the business understanding step, leading to more effective solutions.

## 2.3 Challenges

However, the CRISP-DM model does not clarify the roles of various stakeholders, limiting the value transmission between data science and business. For instance, the roles of stakeholders in the technical process are not clearly defined, potentially resulting in misunderstandings and inadequate communication. Furthermore, the model has not been frequently updated since its development in 1999, so it does not account for modern big data technologies and services. This limitation potentially reduces its relevance and effectiveness in contemporary data science projects.

## 3     The Proposed DataPro Framework

DataPro is a generic framework for data-driven recommendations, and it is an update and extension of the CRISP-DM model. Firstly, to provide a more comprehensive and logical arrangement of project execution activities, two additional phases, "technical understating" and "implementation", are included. Secondly, the roles of data scientists and stakeholders are clearly defined and clarified in this model, facilitating the establishment of effective communication mechanisms to avoid misinformation. Thirdly, the eight phases are classified into five groups to better explain the objectives and activities of each phase.

This section includes two parts: Section 3.1 introduces the general framework of the DataPro model; The details of the activities in each phase and the iteration conditions are discussed in Section 3.2.

### 3.1    Overview

The proposed DataPro framework is demonstrated in Fig. 2. The gray cycle outlines the general explanation of the technical steps, which includes "Define", "Refine", "Analyze", "Delivery" and "Utilize". In the "Refine" and "Analyze" stages, the "Validate" and "Verify" processes are applied for the necessary iterations. The eight blocks inside the circle correspond to the technical phases, with color-coding to indicate the participants: orange represents the stakeholders, while blue denotes data scientists. The black arrows linking the blocks indicate the information flow between the two phases.

The proposed model extends the CRISP-DM framework by incorporating the "technical understanding" and "implementation" phases. The technical understanding phase converts business objectives into data science terminology. This translation facilitates data scientists' comprehension of the business requirements, ensuring that their analytical efforts are precisely aligned with organizational goals. The implementation phase encompasses the practical application of the developed or selected models to real-world scenarios, including model application, parameter fine-tuning, system integration, and so on. This phase is crucial as it transitions the project from theoretical analysis to practical application, demonstrating the real-world value of the data science efforts.

Furthermore, the roles of distinct participants are highlighted in Fig. 2. The business understanding phase is carried out exclusively by the application partners, who are responsible for defining the business objectives of the project. In contrast, the technical understanding, evaluation, and deployment phases necessitate the active involvement of both data scientists and stakeholders. In these three phases, the information is communicated and exchanged between data science and business domains. The collaborative nature of these phases underscores the importance of integrating business insights with technical expertise to achieve successful project outcomes. The subsequent phases are comprised solely of data science tasks.

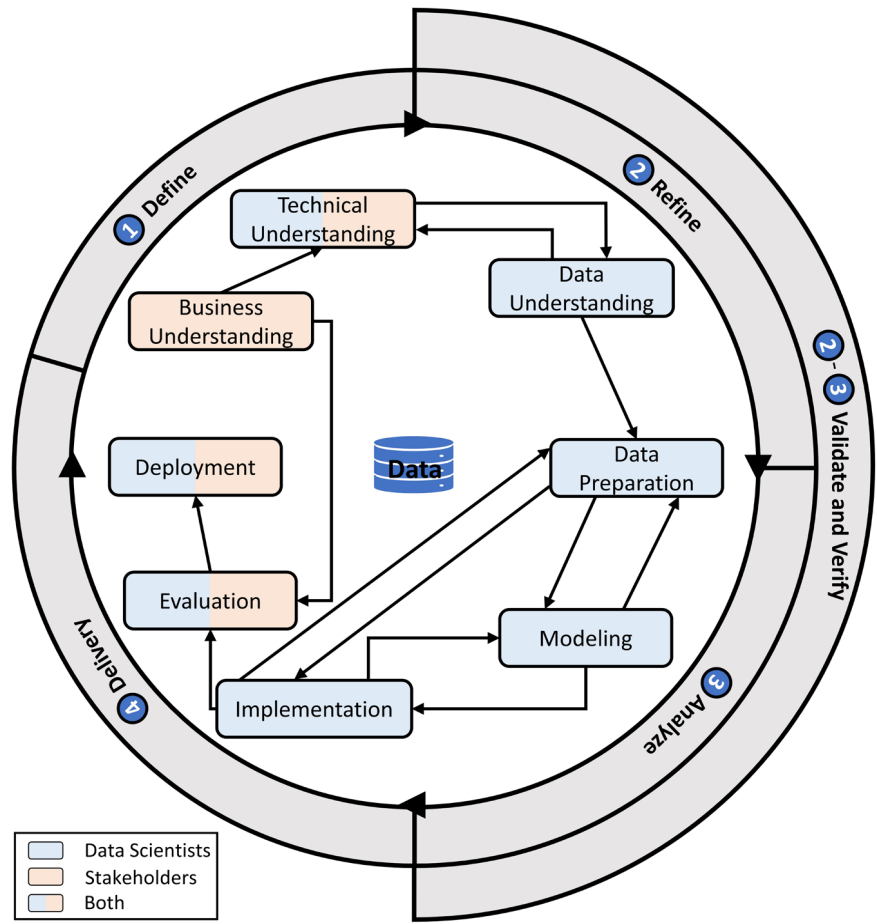

**Fig. 2.** The DataPro process model. The gray cycle outlines the general explanation of the technical steps. The eight blocks inside the circle correspond to the technical phases, with color-coding to indicate the participants as the legend shows. The black arrows linking the blocks indicate the information flow between the two phases.

The definition and activities of each phase, the detailed role of the participants, and the conditions for the iterations are discussed in detail in Section 3.2.

### 3.2    DataPro Framework Elements

**Business Understanding.** The business understanding phase delineates the project objectives from the business perspective. The participants are only the stakeholders, ensuring that the project objectives are aligned with business priorities and strategic goals. The documentation produced in this phase involves the background information regarding the current business situation, specific business objectives, and evaluation criteria from a business perspective [5].

The tool for business understanding is the Business Model Canvas (BMC), a strategic management tool that provides a visual framework for describing, analyzing, and designing business models [13-14]. The BMC comprises nine key components, including key partners, key activities, key resources, value propositions, customer relationships, channels, customer segments, cost structure, and revenue streams. By utilizing the BMC, project participants can gain a comprehensive understanding of the business environment and strategically align their operations to achieve the objectives effectively. Table 1 generalizes the participants, input and output information, and methodology of the business understanding phase.

Table 1. General aspects of business understanding

| Category | Details |
| --- | --- |
| Participants | • Stakeholders |
| Input Information | • N/A |
| Output Information | • BMC results |
| Methodology | • BMC |

**Technical Understanding.** The technical understanding phase is where business objectives and evaluation criteria are translated into a data science framework. This phase involves collaboration between stakeholders and data scientists. Data scientists focus on converting the business goals into technical terms, while stakeholders ensure that the translated objectives and evaluation criteria remain consistent with the original business intents.

In this phase, the technical understanding of a data science project is developed by defining the input variables, output variables, constraint formulas, and quantitative evaluation criteria. Input variables refer to the data that will be used in the analysis, while output variables are the results that the project aims to predict or explain. Constraint formulas establish the boundaries and limitations within which the data science models must operate. Finally, quantitative evaluation criteria are set to measure the effectiveness and accuracy of the data science solutions, ensuring they meet the predefined business objectives. This translation process ensures that technical efforts are aligned with business goals, facilitating the integration of data science solutions into the broader business strategy.

The business understanding and technical understanding phases are both crucial components of the "Define" stage. In this stage, these two phases work together to establish clear and precise objectives and evaluation criteria. This collaboration ensures that business goals are aligned with technical capabilities, facilitating a comprehensive approach that addresses both business and data science challenges in the project.

Table 2 lists the general aspects of the technical understanding phase.

**Table 2.** General aspects of technical understanding

| Category | Details |
| --- | --- |
| Participants | - Data Scientists<br>- Stakeholders |
| Input Information | - BMA results from business understanding<br>- Data quality information from data understanding |
| Output Information | - Input variables, output variables, constraint formulas, and quantitative evaluation criteria of a data science project |
| Methodology | - N/A |

**Data Understanding.** The data understanding phase involves data characteristic understanding and data quality evaluation. This phase aims to identify and mitigate any potential issues during the subsequent data preparation phase. By thoroughly understanding the data's attributes, distribution, and potential anomalies, data scientists can preemptively address problems that might otherwise impede progress. To ensure a comprehensive understanding of data characteristics, the 6Vs model can be adopted, which includes six dimensions: volume, variety, velocity, veracity, value, and variability. This model provides a structured approach to examining the nature of data, facilitating more effective data management and utilization. When conducting data quality evaluations, it is essential to assess several key dimensions to ensure the integrity and utility of the data. These dimensions include completeness, timeliness, accuracy, consistency, validity, interpretability, relevancy, and uniqueness [15].

From the data understanding phase to the implementation phase, the involvement is limited to data scientists, as these tasks require specialized technical expertise and do not rely on business understanding.

There exists a feedback loop between data understanding and technical understanding, where the data quality significantly influences the feasibility and success of technical objectives. This interdependency implies that technical goals often cannot be realized without data quality issues. Consequently, it is imperative to adjust both the objectives and evaluation criteria based on data quality challenges, such as incomplete or invalid data.

Table 3 demonstrates the general aspects of the data understanding phase.

**Table 3.** General aspects of data understanding

| Category | Details |
| --- | --- |
| Participants | - Data Scientists |
| Input Information | - Outputs from technical understanding |
| Output Information | - Data quality understanding results (delivered to both technical understanding and data preparation) |
| Methodology | • 6V data understanding framework [12] |

**Data Preparation.** The data preparation data phase preprocesses the datasets based on the insights from the data understanding phase. Key activities in this phase include data cleaning, data transformation, and data integration [16]. By performing these activities, the data preparation phase ensures that datasets are clean, accurate, and well-structured, thereby enhancing the effectiveness and efficiency of the subsequent modeling phases.

Data cleaning involves identifying and correcting errors and inconsistencies in the data, such as handling missing values, correcting inaccuracies, processing outliers, and removing duplicates. Furthermore, feature selection is another important task in data cleaning, which involves identifying and retaining the most relevant variables for analysis, thereby enhancing the quality and performance of the data processing model in the next phase.

Data transformation methods convert data into a suitable format for analysis. Tasks include normalization, aggregation, and encoding categorical variables, ensuring the data is standardized and comparable.

Data integration methods combine data generated from various sources to create a unified dataset. Tasks include merging datasets, reconciling data discrepancies, and ensuring consistency across the integrated data.

The data understanding and data preparation phases encompass the "Refine" stage, which focuses on improving the data quality for modeling. These phases ensure that the data is thoroughly understood and preprocessed, addressing inconsistencies and errors, and integrating data from various sources.

Table 4 presents the general aspects of the data preparation phase.

**Table 4.** General aspects of data preparation

| Category | Details |
|---|---|
| Participants | • Data Scientists |
| Input Information | • Synthetic and/or real-world datasets |
| | • Outputs from data understanding |
| | • Feedback from modeling regarding new issues and challenges in modeling experiments |
| | • Feedback from implementation regarding issues and challenges of real-world data in implementation experiments |
| Output Information | • Preprocessed datasets |
| Methodology | • Data preprocessing methods, such as data cleaning, data transformation, and data integration |

**Modeling.** In the modeling phase, various models are run using default parameters initially and the model with best performance is selected. Following this, fine-tuning of the parameters of the selected model is performed to optimize the model's performance. This phase encompasses a range of models, including statistical methods such as causal inference and hypothesis testing, as well as machine learning and deep learning algorithms for tasks like prediction, classification, and clustering.

If necessary, the process may revert to the data preparation phase for additional data manipulations required by the chosen model. Therefore, there is a constant feedback loop between data preparation and modeling. As new issues and challenges are identified during modeling experiments, data preprocessing strategies must be adjusted to better align the data with the requirements of the model, thereby improving its performance and accuracy.

Thus, the data preparation phase works together with the modeling phase within the "Analyze" stage. This collaboration is essential for continuous refinement and improvement of the data, ensuring that it is of the highest quality and suitability for modeling. By maintaining this iterative approach, organizations can better identify and rectify issues, ultimately leading to more reliable and robust models that provide valuable insights and support informed recommendations.

Table 5 demonstrates the general aspects of the modeling phase.

**Table 5.** General aspects of modeling

| Category | Details |
| --- | --- |
| Participants | - Data Scientists |
| Input Information | - Preprocessed datasets |
|  | - Feedback from implementation regarding model's performance |
| Output Information | - Selected models for implementation |
|  | - Parameter selection procedures for models in implementation |
|  | - Feedback to data preparation regarding data-related issues and challenges in modeling experiments |
| Methodology | - Data analytic methods, such as statistical tests and machine learning algorithms |
|  | - Quantitative evaluation methods |

**Implementation.** The implementation phase applies the models developed in the modeling phase to a real-world environment. Key activities include integrating the model into real-world systems, continuously monitoring its performance on diverse real-world datasets, and making necessary adjustments to ensure robustness and reliability. This phase also involves scaling the model to handle large volumes of real-world data and ensuring it operates efficiently under various usage conditions.

Additionally, the implementation phase relies on preprocessed datasets received from the data preparation phase for testing and validation. Moreover, continuous feedback loops between the modeling and implementation phases are essential to maintain and improve the model's performance, addressing any issues that arise in the application environment. Overall, successful implementation ensures that the model delivers actionable and reliable insights that can be effectively used for recommendation.

Table 6 lists the general aspects of the implementation phase.

Table 6. General aspects of implementation

| Category | Details |
| --- | --- |
| Participants | - Data Scientists |
| Input Information | - Prepared real-world datasets |
|  | - Selected models for implementation |
|  | - Selected parameters for models in implementation |
| Output Information | - Integrated model in real-world systems |
|  | - Feedback to data preparation regarding real-world data-related issues and challenges |
|  | - Feedback to modeling regarding issues that arise in the application environment |
| Methodology | - Model integration and validation in the real-world environment |
|  | - Quantitative evaluation methods |

**Evaluation.** In the modeling and implementation phases, the constructed models have been assessed based on established technical evaluation criteria, ensuring that the models are technically sound and effective. Subsequently, in this evaluation phase, it is crucial to assess the outcomes using the evaluation criteria defined in the business understanding phase. This dual-layered evaluation approach guarantees that the technical accuracy is complemented by alignment with business objectives.

In the evaluation phase, the technical results need to be translated into operational recommendations. This translation is essential to bridge the gap between technical analytics and practical business insights, making the findings accessible and actionable for stakeholders, who are primarily concerned with the operational and strategic implications.

By converting technical results into understandable recommendations, the business participants can effectively utilize the insights to drive decision-making processes. This collaborative approach not only enhances the overall effectiveness of the modeling process but also ensures that the technical and business perspectives are harmoniously integrated, fostering a comprehensive understanding and application of the results.

Table 7 demonstrates the general aspects of the evaluation phase.

Table 7. General aspects of evaluation

| Category | Details |
| --- | --- |
| Participants | - Data Scientists |
|  | - Stakeholders |
| Input Information | - Evaluation criteria from business understanding |
|  | - Integrated model and results from implementation |
| Output Information | - Application and decision-making recommendations for stakeholders |
|  | - Evaluation results |
| Methodology | - Qualitative and quantitative evaluation methods |

**Deployment.** The deployment phase represents the final step of the data science project, which delivers the values to various stakeholders and concludes the project. The deployment phase includes two activities: planning and monitoring the deployment of results and producing a final report.

Planning and monitoring ensure that the deployment process is smooth, efficient, and aligned with stakeholders' expectations. This involves continuous oversight to address any issues that arise during implementation. Furthermore, the final report documents the project outcomes, methodologies, and insights, encompassing the models, technical results, and business recommendations. This report also includes a project review to assess the overall performance and identify lessons learned.

The implementation, evaluation, and deployment phases consist of the "Delivery" stage, ensuring that models are applicable in real-world scenarios while delivering business value propositions. This stage is crucial for validating the practical applicability and alignment of models with operational goals and strategic objectives.

Table 8 presents the general aspects of the deployment phase.

**Table 8.** General aspects of deployment

| Category | Details |
|---|---|
| Participants | - Data Scientists<br>- Stakeholders |
| Input Information | - Evaluation results and recommendations from the evaluation phase |
| Output Information | - Final report for stakeholders and data scientists |
| Activity | - Planning and monitoring the deployment of results<br>- Producing a final report |

## 4    Case Study

### 4.1    Project Description

This case study, part of a funded national research project, demonstrates the application of the DataPro framework in a real-world industrial setting, focusing on reducing energy consumption and greenhouse gas emissions in the public transportation sector. Modern transport remains dependent on internal combustion engines that generally run on fossil fuels, and the transportation sector accounts for more than a third of $CO_2$ emissions in end-use sectors [17-19].

This project aims to reduce energy consumption and greenhouse gas emissions in the public transportation sector, specifically in a Danish public transportation system. The project involves collaboration between a transportation company, a software company, and a university. The software company is responsible for sensor implementation, data collection, and data management, while the university applies the DataPro framework to conduct the data understanding and analytic process and provide recommendations to the end user.

### 4.2   DataPro Application

The DataPro framework's application is methodically executed through several phases, each contributing to the overarching goal of energy and emissions reduction:

**Business Understanding.** The primary business objectives, including reducing $CO_2$ emissions, lowering energy costs, and enhancing energy efficiency, are identified and validated using the Business Model Canvas (BMC) [14]. This tool helps in thoroughly understanding the business context and demands from the end-user, as illustrated in Table 9.

Table 9. The BMC of the industrial case study.

| BMC components | Details |
|---|---|
| Key Partners | • Data scientists<br>• Transportation end-users<br>• Software developers |
| Key Activities | • Big data analysis<br>• AI<br>• Statistical analysis |
| Key Resources | • Skilled professionals<br>• Software and platform |
| Value Propositions | • $CO_2$ emission reduction<br>• Factors influencing energy consumption |
| Customer Relationships | • Consultation<br>• Maintenance service |
| Channels | • Online platforms |
| Customer Segments | • Production data management<br>• R&D |
| Cost Structure | • Operational costs |
| Revenue Streams | • Financial inflow |

**Technical Understanding Phase.** Information from the BMC is translated into a technical framework. This involves defining input data, expected outputs, evaluation metrics, and operational constraints. Necessary communication is conducted between data scientists and end-users through the channel shown in Table 1. Key elements of technical understanding are detailed in Table 10.

Table 10. Technical understanding results.

| Categories | Elements |
|---|---|
| Input | • Sensor data from the buses |
| Output | • Best practice operation<br>• Factors influencing fuel consumption |
| Evaluation metrics | • Evaluation metrics for clustering |
| Constraints | • Statistical evaluation of influencing factors<br>• Drivers' natural behavior in driving<br>• Safety regulations and laws in driving |

**Data Understanding Phase**: Data scientists analyze the characteristics of the data provided by the software company and the end-user using the 6Vs model [12], addressing issues like missing values and duplicated timestamps to ensure high data quality for analysis. The data understanding phase results verify that technical objectives can be achieved with the current data.

**Data Preparation Phase**: Features with 99% missing values are removed, and the remaining missing values are imputed based on time series distributions. Duplicated timestamps are integrated using established methodologies introduced in [12]. Then a new feature termed "fuel efficiency" is introduced by combining two existing features through a specific calculation. Subsequently, the characteristic features are extracted because the dataset of each trip is a multivariate time series, whereas the fuel efficiency is represented by a single value. The refined datasets are prepared for the modeling and implementation phases.

**Modeling Phase**: The Gaussian mixture model (GMM) clustering algorithm is employed to identify the fuel-consumption best-practice group. Three evaluation indexes, including the Silhouette index (SI), Calinski-Harabasz index (CHI), and Davies-Bouldin index (DBI), are integrated to select the optimal number of clusters and assess the clustering performance. Next, the two-sample Kolmogorov-Smirnov test (KS test) is utilized to analyze whether individual driving behavior and route conditions significantly impact fuel consumption. The p-value is the evaluation criteria. The prepared datasets are of sufficient quality for modeling; thus, no further data preprocessing is required for implementation.

**Implementation Phase**: A data processing pipeline is implemented using the Visual Studio Code platform (version 1.85.2) with all code written in Python (version 3.10.9). Four clusters are identified based on the evaluation indexes, including the extremely fuel-efficiency group, normal fuel-efficiency group, low fuel-efficiency group, and extremely low fuel-efficiency group. Next, the KS test indicates that the fuel efficiency of distinct drivers and routes varies significantly. The model's effectiveness is validated, negating the need for further iterations.

**Evaluation Phase:** The clustering results are assessed through three evaluation indexes. The variance in fuel efficiency of distinct drivers and routes highlights the significant influence of individual driving habits and specific route characteristics, underscoring the need for personalized approaches in improving fuel efficiency and optimizing route planning for better performance outcomes.

**Deployment Phase**: The project concludes with three comprehensive reports:
- Data Science Perspective: Summarizing data preprocessing, model selection, and evaluation for future projects, and highlighting the direction of the following project: causal inference of fuel efficiency in each group.
- End-user Recommendations: Detailing driving behavior and route planning to improve fuel efficiency.
- Software Company Guidance: Providing insights into improving data collection methods and coping with associated risks.

## 5      Qualitative Evaluation of the DataPro Model

### 5.1     Evaluation Criteria

The qualitative evaluation of data understanding and processing procedures can be assessed through the following four key dimensions.

1. Good Data Preparation (GoodDP): This criterion emphasizes comprehensive data preprocessing to prepare high-quality data for subsequent reliable data analytics.
2. Accurate Business Understanding (AccuBU): A clear and precise understanding of business objectives ensures that the data analytic efforts are aligned with the strategic goals.
3. Flexible Iteration (FlexIter): The ability to iterate flexibly through necessary steps of the data analytic processes is vital. Revisiting and refining various stages based on the findings and feedback obtained throughout the process allows for adaptive adjustments, improving the effectiveness and accuracy of the data processing results.
4. Clear Responsibility (ClearRes): Defining and clarifying the responsibilities of various stakeholders at each step of the data understanding and processing procedure ensures that all parties involved understand their roles and duties, which fosters accountability and streamlines collaboration.

The score for each criterion is selected from {0.3, 0.7, 1.0}. A score of 0.3 denotes that the process weakly matches the definition of the criterion. A score of 0.7 indicates that the process matches most aspects of the definition. A score of 1.0 signifies that the process fully matches the definition.

### 5.2     Baseline Models

The baseline models compared to the DataPro process include KDD, SEMMA, and CRISP-DM. The CRISP-DM model has been introduced in Section 2; therefore, KDD and SEMMA are briefly introduced in this section.

The KDD (Knowledge Discovery in Databases) process [20] involves several stages: data selection, data preprocessing, data transformation, data mining, interpretation/evaluation, and knowledge presentation. This process focuses on transforming raw data into useful knowledge, emphasizing the iterative nature and the need for continuous evaluation and refinement in each step.

The SEMMA framework, proposed by the SAS Institute [21], encompasses sampling, exploring, modifying, modeling, and assessing. This model emphasizes the technical aspects of data science projects, particularly the modeling phase.

### 5.3     Evaluation Results and Discussion

Table 11 demonstrates the evaluation results of the four data science procedures.

Table 11. Qualitative evaluation scores.

| Models   | GoodDP | AccuBU | FlexIter | ClearRes |
|----------|--------|--------|----------|----------|
| KDD      | **1.0** | 0.3   | **1.0**  | 0.3      |
| SEMMA    | 0.3    | 0.3    | 0.3      | 0.3      |
| CRISP-DM | **1.0** | 0.7   | 0.7      | 0.3      |
| DataPro  | **1.0** | **1.0** | 0.7    | **1.0**  |

The KDD, CRISP-DM, and DataPro methodologies include specific phases dedicated to data preprocessing or data preparation. In contrast, the SEMMA framework lacks such a procedure, utilizing the samples directly without additional preprocessing steps.

Both CRISP-DM and DataPro incorporate a "business understanding" phase. DataPro further includes a subsequent "technical understanding" phase, which translates business objectives into technical expressions to ensure accurate comprehension by data scientists. In contrast, KDD and SEMMA primarily focus on the technical aspects of the data processing procedure.

The KDD, CRISP-DM, and DataPro methodologies define iterative processes within their frameworks. KDD allows for iterations between any two steps based on the data science nature, offering greater flexibility. In contrast, CRISP-DM and DataPro specify particular iterative paths. SEMMA, however, does not clearly define the flow of iterations, potentially leading to ambiguities.

Moreover, DataPro uniquely provides clear definitions of the tasks and responsibilities assigned to data scientists and stakeholders.

In summary, DataPro is the most effective and comprehensive model based on the four evaluation criteria.

## 6     Discussions and Conclusion

This paper proposes the DataPro framework, an extension of CRISP-DM, to enhance data-driven recommendations in data science projects. This framework introduces two extra key phases: the "technical understanding" phase and the "implementation" phase, aiming to strengthen the connection between data scientists and stakeholders. The technical understanding phase bridges the gap between business demands and technical requirements, ensuring data scientists accurately comprehend business goals. The implementation phase ensures theoretical models are effectively translated into actionable business solutions. Additionally, the framework outlines clear roles and responsibilities for various participants in each phase, improving management and communication. By providing clarity on these aspects, the framework facilitates smoother and more efficient data science project progression, fostering enhanced collaboration among all involved parties.

The case study in the transportation sector demonstrated the practical application of the DataPro framework. This project aimed to reduce energy consumption and greenhouse gas emissions in the Danish public transportation system by following the DataPro framework to identify key business objectives, translate them into technical requirements, develop actionable models, and provide data-driven insights. Key findings include identifying four fuel efficiency groups and the significant impact of driving behavior and route characteristics. These insights were validated through rigorous evaluation methods, highlighting the framework's effectiveness in delivering practical, data-driven recommendations. More data science projects will be undertaken using the DataPro procedure to validate the proposed framework's applicability across various contexts.

The qualitative evaluation of the DataPro framework against KDD, SEMMA, and CRISP-DM showcased its superiority. This comprehensive evaluation highlights DataPro's strengths in providing a structured and iterative approach to data science projects, ensuring both technical excellence and business relevance. However, future work should aim to update the evaluation metrics to capture the practical implementation and scalability aspects better. This enhancement would provide a more comprehensive analysis, ensuring the results are applicable and beneficial in real-world scenarios.

Furthermore, the framework's implementation in this study is limited to a single domain. Future research should focus on validating the DataPro framework across various industries to ensure its versatility and robustness. Additional data science projects can provide deeper insights into the framework's adaptability and effectiveness in different contexts. Furthermore, future work should delve into providing clearer technical application details, including opportunities to conduct iterations. Specifically, the iterative aspects of the framework should be explored in detail to understand how they can be optimized for different project scales and complexities.

Moreover, it is crucial to investigate the integration of emerging technologies such as edge computing, IoT, and advanced machine learning algorithms within the DataPro framework. For instance, IoT service companies could assume responsibility for data collection and management. Moreover, employing more interpretable machine learning algorithms is crucial to enhance the explainability of model implementation. These technologies have the potential to enhance data processing capabilities, improve data security, and offer more sophisticated analytical tools, thereby broadening the scope and impact of data-driven projects.

**Acknowledgments.** This paper is part of the project "Driver Coach", funded by the Danish funding agency, the Danish Energy Technology Development, and Demonstration (EUDP) program, Denmark (Case no. 64020-2034), the IEA IETS Task XVIII: Digitalization, Artificial Intelligence, and Related Technologies for Energy Efficiency and GHG Emissions Reduction in Industry funded by EUDP (Case no. 134-21010), and the project "Data-driven best-practice for energy-efficient operation of industrial processes - A system integration approach to reduce the $CO_2$ emissions of industrial processes" funded by EUDP (Case no. 64021-2108).

**Disclosure of Interests.** The authors have no competing interests to declare that are relevant to the content of this article.